\documentclass[%
 reprint,
 amsmath,amssymb,
 aps,
]{revtex4-2}

\usepackage{graphicx}% Include figure files
\usepackage{dcolumn}% Align table columns on decimal point
\usepackage{bm}% bold math
\usepackage{xcolor}
\begin{document}

\preprint{APS/123-QED}

\title{Oscillatory and dissipative dynamics of complex probability in non-equilibrium stochastic processes}% Force line breaks with \\

\author{Anwesha Chattopadhyay}
 \affiliation{Department of Physics, School of Mathematical Sciences,
Ramakrishna Mission Vivekananda Educational and Research Institute, Belur, Howrah 711202, India.}%Lines break automatically or can be forced with \\
\date{\today}% It is always \today, today,
\begin{abstract}
For a Markov and stationary stochastic process described by the well-known classical master equation, we introduce complex transition rates instead of real transition rates to study the pre-thermal oscillatory behaviour in complex probabilities. Further, for purely imaginary transition rates we obtain persistent infinitely long lived oscillations in complex probability whose nature depends on the dimensionality of the state space. We also take a peek into cases where we perturb the relaxation matrix for a  dichotomous process with an oscillatory drive where the relative sign of the angular frequency of the drive decides whether there will be dissipation in the complex probability or not. 
\end{abstract}

\maketitle

%\tableofcontents

\section{Introduction}

The classical master equation for a Markov and stationary stochastic process defines a relaxation matrix in terms of the transition rates from one state to the other~\cite{Bala}. Such transition rates are real. In addition, the sum of the column elements of the relaxation matrix vanish which means it has a zero determinant and hence one of the eigenvalues is necessarily zero which is an essential condition for the system to equilibriate. The matrix is generally non-symmetric and the eigenvalues are in general complex. However, the reality of the trace ensures that the other eigenvalues (apart from zero) occur in complex conjugate pairs. Another essential condition for equilibriation is that the real part of the eigenvalues must be negative which would give exponentially decaying contribution in time from all non-trivial eigenvalues and hence a steady state behaviour in the long time limit. Gershgorin disc theorem~\cite{Bala} provides a way of seeing that this is really the case for an equilibriating system.

However, in this paper, we firstly, explore the case of complex transition rates instead of real transition rates, as is common in literature. We retain all other aspects of the relaxation matrix in the sense that the sum of the columns are still zero. However, the trace is no longer real and so the eigenvalues do not necessarily occur in conjugate pairs. The complexity of the transition rates introduces the concept of complex conditional probabilities~\cite{ComplexProb1,ComplexProb2}. We observe interesting pre-thermal oscillations in complex probability in this case when we start with a general initial state. We explore the dyanamics initially for a dichotomous process and then we extend it for the case of an $N$-state stochastic system. There is an unique state having equal support on all the states of the state space which does not show any pre-thermal dynamics and remain frozen to a single value. However, the total complex probability for all states is conserved. Next we venture further, as we take the example of the dichotomous process again and drive it with an oscillatory field. As we know the master equation for the rate of change in conditional probability  for a particular state has two terms: one is the loss term which gives the rate at which conditional probability is being lost from the specific state to other states in the state space and the other is the gain term which gives the rate at which conditional probability is being gained from the other states of the state space. Coupling the usual relaxation matrix with a diagonal matrix with equal imaginary terms physically means that we are introducing with the coefficient of loss term, an extra term which returns the state to itself with a particular angular frequency. This means we drive both the states of the system back and froth at the same rate so that the probabilities remain the same as that of the original problem modulo some phase which is oscillatory in time. The total probability is also unity modulo a phase which is oscillatory in time. So, the inclusion of a symmetric driving field in a two state process does not alter the physics of the problem in any substantial way. Similarly, instead of altering the loss term if we change the rate of gain by a symmetric imaginary term, then the probability dynamics is the same as that of the problem of complex transition rates modulo a phase factor oscillating in time. On the other hand if we couple the relaxation matrix with a driving field which rotates the complex probability vector of one state in the clockwise and other in the anti-clockwise sense, non-trivial dynamics is observed. More importantly, the total probability dissipates and in the long time limit the system disintegrates or melts so that the probability to be in each state is identically zero.

Even though all the analysis we present in this paper is from  a classical point of view, this analysis can be extended to quantum master equations where the transition rates are derived from the Fermi-Golden rule~\cite{FermiGolden}. Taking complex transition rates in the classical master equation is analogous to coupling the transition rates already predicted by Fermi-Golden rule with an imaginary parameter. {
The concept of complex probability in a classical purview, at first may seem physically unrealistic. It is true that probability should be a positive real number which is measurable in experiments. However, the concept of a complex probability phasor which appears solely due to the introduction of complex transition rates has a one to one correspondence with probability amplitude in quantum mechanical systems. More precisely, if the quantum master equation is solved with complex transition rates, the probability amplitude which is inherently complex would show dynamics reminiscent of what the classical complex probability phasor is doing here. The complexity of the classical probability is then just an abstract artifact which arises when we take the classical limit of the quantum probability amplitude which itself is inherently complex. However, as a first step in this endeavor, we would concentrate only on the classical master equation in this paper.} The interest in exploring the cases mentioned above stems from the recent interest in the statistical physics community in the topic of weak ergodicity violation in isolated quantum systems~\cite{Isolated}. In isolated quantum systems, a subsystem is thermalized by the complementary of that subsystem which acts as it's bath. So, local operators which have support on this subsystem show thermalizing behaviour for an ergodic system. However, it has been observed that there are some special initial states for which such a local operator avoid thermalization by showing persistent long time oscillations. These special states show large overlap with certain algebraic number of  eigenstates which show sub-volume entanglement entropy and are dubbed as ``quantum scars"~\cite{Scar1,Scar2,Scar3}. As in this system, there is no notion of external bath in the classical system that we explore. However, analogously we can say, that complementary of each state in the state space is acting as it's bath and eventually thermalizing the system. Taking purely imaginary transition rates, induces persistent oscillations in the complex probabilities which holds some interest in understanding the microscopic origin of these quantum scar states.{ This classical analysis therefore hints that it is highly probable that the persistent fidelity oscillations of dynamical observables in some initial states which have large overlap with quantum scar states arise because of some intrinsic complex transition rate in the system. The imaginary part of the complex transition rate signifies rebound of the system from that state to itself in a to and fro manner, thus giving rise to oscillatory dynamics.  There may be virtual states existing in the system which assist the system in getting scattered back to that state.}. Further, these and related phenomena can be observed in ultracold platforms with two level systems like Rydberg atoms~\cite{Rydberg1,Rydberg2}. {We know that the transition rates can be tuned by lasers in case of Rydberg atoms. As emphasized earlier, if we look at the loss term of the master equation, the imaginary term in the transition rate returns the system in some initial state to itself where as the real term in the transition rate takes the system from that state to other states of the state space. Therefore, introducing an impurity state which reflects the two level system say from the ground state to itself while it tries to make a jump to the excited state, mimics a complex transition rate in a two level system.}

{Another point to ponder about is the role of environment and temperature in our analysis. While our current model does not explicitly include environmental interactions, we agree that it is an inevitable aspect of complex systems. Moreover, stochastic fluctuations like noise~\cite{noise1,noise2,noise3,noise4,noise5,noise6,noise7}, in both classical and quantum setups, can be used in constructive ways to enhance stability, coherence, or even induce new dynamical regimes in non-equilibrium systems. An interesting phenomenon is the noise-enhanced stability (NES)~\cite{NES1,NES2, NES3}, where noise increases the residence time in a metastable state and is a stochastically driven process. This has conceptual similarity with the persistent oscillations observed in our model since both refer to the stability of non-equilibrium states. However, the persistent oscillations observed in our case arise from deterministic intrinsic bounce-back dynamics modeled by imaginary transition rates rather than stochasticity. Stochasticity in our model enters through the real part of the transition rate which is a function of the temperature of the coupling thermal bath and hence it controls the longevity of the pre-thermal oscillatory regime. Stochasticity, is however, not a pre-requisite for observing persistent oscillations between the states. Effect of temperature is encoded in the real part of the transition rate. However, if we explicitly introduce non-thermal noises in our model, something akin to NES  may be observed if there is noise aided stabilization of oscillations. This counter-intuitive expectation is not very drastic as it has been recently observed that correlated noise can be effectively harnessed as a
valuable resource for enabling controlled entanglement generation, specifically achieving
enhanced and long-lived entanglement~\cite{npj}. However, higher intensities of noise may suppress the oscillations.}

{ In section II, we give details of the classical master equation. In the subsection A, we study the case of real transition rates and in subsection B, we study the case for complex transition rates. In subsection C, we study the case of probability dissipation. Finally, in section III, we conclude.}

\section {The Master equation}

The differential equation satisfied by the conditional probabilities of a stationary and Markovian random process is obtained from the Chapman–Kolmogorov equation~\cite{Bala},

\begin{equation}{\label{e:Eqn1}}
    P(k,t|j) = \sum_{l=1}^{N}P(k,t-t' |l) P(l, t'| j)
\end{equation}

where $t' \in (0,t)$. This equation says that the probability of propagation from the initial state $j$ to final state $k$ in time $t$ is the product of probabilities of propagation form $j$ to some intermediate state $l$ in time $t'$ and the probability of propagation from $l$ to $k$ in the remaining time interval $t-t'$. This is summed over all intermediate states $l$.

To linearize this equation, the concept of transition rates $w(k|j)$ is introduced which is the probability of transition per unit time from state $j$ to $k$ for $j \neq k$. Mathematically, the conditional probability $P(k,\delta t|j)$ is proportional to the infinitesmal time interval $\delta t$ with the constant of proportionality being the transition rate $w(k|j)$ such that,

\begin{equation}{\label{e:Eqn2}}
    P(k,\delta t|j)=w(k|j)\delta t 
\end{equation}

where $k \neq j$. Using \eqref{e:Eqn1},\eqref{e:Eqn2} we can derive a set of linear but coupled differential equations satisfied by the conditional probabilities $\{P(k,t|j)\}$,

\begin{equation}
    \dfrac{d}{dt}P(k,t|j)=\sum_{\substack{{l=1}\\{l\neq k}}}^{N} w(k|l)P(l,t|j)-w(l|k)P(k,t|j).
\end{equation}

The first term is the gain term which gives the rate at which conditional probability is gained in state $k$ from all other states in the state space while the second term is the loss term indicating the rate at which conditional probability is lost from the state $k$ to other states of the state space. This equation can be reframed into a matrix equation,

\begin{equation}
    \dfrac{d}{dt}\mathcal{P}=\mathcal{W}\mathcal{P} \label {e:Eqn3}
\end{equation}

where $\mathcal{P}$ is a $N \times 1$ matrix of conditional probabilities of $N$ states at time $t$ subject to the condition that initially the system is in some discrete state or a linear combination of such states. $\mathcal{W}$ is a $N \times N$ matrix known as the relaxation matrix which in the usual cases determine the relaxation of the system to a stationary distribution. The diagonal elements of $\mathcal{W}$ are $\mathcal{W}_{kk}=-\sum_{\substack{{l=1}\\{l\neq k}}}^{N}w(l|k)$ where as the off-diagonal terms are $\mathcal{W}_{kj}=w(k|j)$. We can solve \eqref{e:Eqn3} in terms of the initial distribution $\mathcal{P}(0)$ as,

\begin{equation}
    \mathcal{P}(t)=e^{\mathcal{W}t}\mathcal{P}(0).\label{e:Eqn4}
\end{equation}

\subsection{Real transition rates}

First we describe the dynamics of conditional probabilities taking the transition rates $w(k|j), k\neq j$ to be real and positive as is common in literature. As we can see from \eqref{e:Eqn4} that the eigenvalues of $\mathcal{W}$ determine the dynamics of $\mathcal{P}(t)$. The sum of the elements of the columns of $\mathcal{W}$ is zero which means it's determinant is also zero and hence zero is necessarily an eigenvalue of $\mathcal{W}$. In addition, $\mathcal{W}$ is in general a non-symmetric matrix and hence it can have complex eigenvalues which must occur in conjugate pairs due to the reality of the trace. Moreover, if the distribution has to reach a stationary state, the real parts of these eigenvalues have to be negative because then they would contribute an exponentially decaying term in time which would disappear in the long time limit leading to the stationary distribution determined by the zero eigenvalue of the relaxation matrix. This is what generally happens for an equilibriating system.

The Gershgorin disc theorem can be used to prove that the real part of the eigenvalues of $\mathcal{W}$ are negative. In the complex plane, if we draw circles with radius $\sum_{\substack{{j=1}\\{j\neq k}}}^{N}\mathcal{W}_{jk}$ centered around $\mathcal{W}_{kk}$ for every column $k$, then all the eigenvalues of $\mathcal{W}$ lie within the union of these circles. If there is a disjoint circle, then it has exactly one eigenvalue. Since the sum of the columns of $\mathcal{W}$ is zero, the radius of the circle is $-\mathcal{W}_{kk}$. The centers of the circles lie on the negative real axis as $\mathcal{W}_{kk}<0$ and since the radius is  $-\mathcal{W}_{kk}$, all of them pass through the origin, telling us that zero is an eigenvalue. And since all the circles are in the second and third quadrant of the complex plane, the real parts of the eigenvalues are necessarily negative. 

\begin{figure}[ht!]
     \begin{center}
\includegraphics[height=3.6cm]{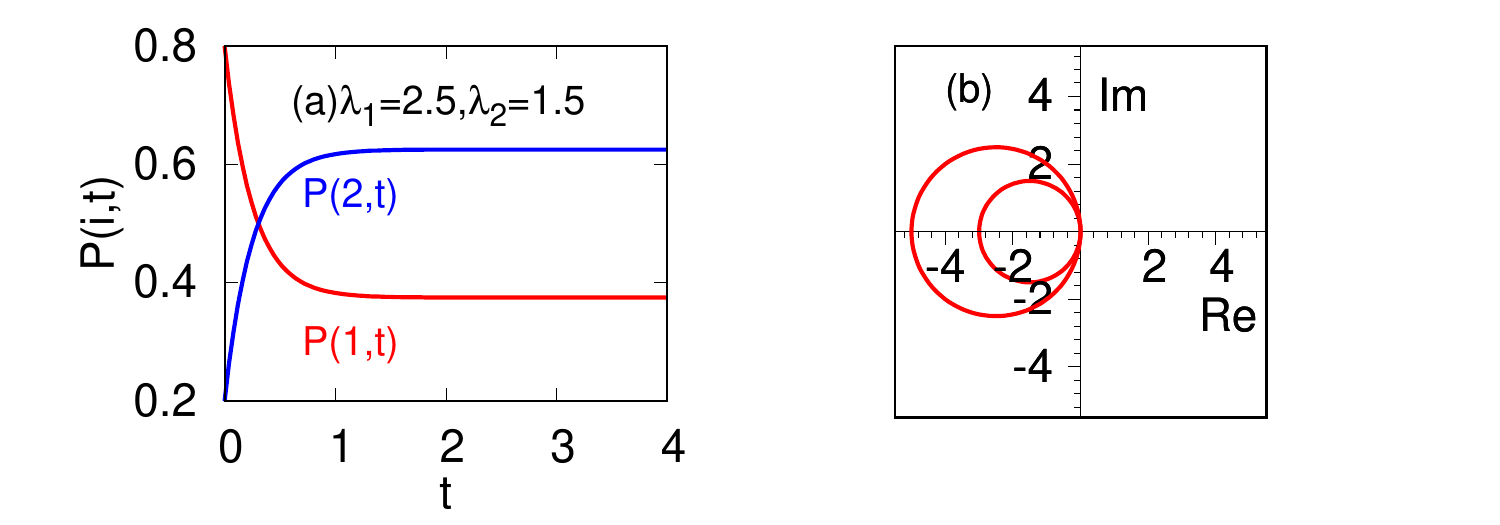}
\end{center}
   \caption{Panel (a) shows that for $\lambda_{1}=2.5, \lambda=1.5$ and mean transition rate $\lambda=2$ in units of inverse time, the conditional probabilities forget about the initial state memory, $P(1,0)=0.8, P(2,0)=0.2$ to equilibriate to their thermal values. Panel (b) shows that for such a choice of $\lambda_{1}$ and $\lambda_{2}$, th Gershgorin discs lie in the second and third quadrant of the complex plane. The eigenvalues of the $\mathcal{W}$ matrix lie in the union of these two circles signifying that the real part of the eigenvalues are always negative. In this case the eigenvalues are 0 and $-2\lambda$.}{\label{fig1}}
  % \label{fig:subfigures}
\end{figure}

Let us take the example of a dichotomous process with real and positive transition rates $w(1|2)=\lambda_2$ and $w(2|1)=\lambda_1$ such that,

\begin{equation}
    \mathcal{W}=\begin{pmatrix}
        -\lambda_{1} & \lambda_2\\
        \lambda_{1} & -\lambda_2
    \end{pmatrix}.
\end{equation}

Exponentiating this $2\times2$ matrix we can find the conditional probabilities at time $t$ starting from some initial distribution say, $\mathcal{P}(0)=\begin{pmatrix}
    a\\
    b
\end{pmatrix}$ such that $a+b=1$ and,

\begin{equation}
\begin{aligned}
        e^{\mathcal{W}t}=&\sum_{n=0}^{\infty}\mathcal{W}^{n}\dfrac{t^{n}}{n!}\\=&\mathbb{I}-\sum_{n=1}^{\infty}\dfrac{(-2\lambda t)^{n}}{n!}\dfrac{\mathcal{W}}{2\lambda}\\=&\mathbb{I}-\dfrac{\mathcal{W}}{2\lambda}(e^{-2\lambda t}-\mathbb{I}).
\end{aligned}
\end{equation}

Here, $\lambda=(\lambda_{1}+\lambda_{2})/2$ is the mean transition rate. Therefore,

\begin{equation}
\begin{aligned}
   & P(1,t)=a-\dfrac{\lambda_{2}b-\lambda_{1}a}{2\lambda}(e^{-2\lambda t}-1)\\
    &  P(2,t)=b-\dfrac{\lambda_{1}a-\lambda_{2}b}{2\lambda}(e^{-2\lambda t}-1)
      \end{aligned}
\end{equation}

such that $P(1,t)+P(2,t)=1$ at all times. Fig~\ref{fig1} shows the conditional probabilities as a function of elapsed time since the system was prepared in an initial state $P(1,0)=0.8, P(2,0)=0.2$ with dynamics governed by transition rates $\lambda_{1}=2.5, \lambda_{2}=1.5$ in units of inverse time. They ultimately reach their equilibrium values where detailed balance is obeyed and the dynamics is frozen thereafter. Also shown are the corresponding Gershgorin circles for this case, the union of which contains the eigenvalues $0,-2\lambda$. 

\subsection{Complex transition rates}

We revisit the two state problem of the previous section, only now with complex transition rates. The $\mathcal{W}$ matrix can now be broken down into two parts: the usual relaxation matrix with real and positive transition rates and a matrix with imaginary components such that the rule of vanishing of the sum of the elements of the columns still holds. The $\mathcal{W}$ matrix is, 

\begin{equation}
        \begin{aligned}
    \mathcal{W}=&\begin{pmatrix}
    -\lambda_{1}-i\omega & \lambda_{2}+i\omega\\
     \lambda_{1}+i\omega & -\lambda_{2}-i\omega\end{pmatrix}\\
    =& \begin{pmatrix}
        -\lambda_{1} & \lambda_2\\
        \lambda_{1} & -\lambda_2
    \end{pmatrix} +     \begin{pmatrix}
        -i\omega & i\omega\\
        i\omega & -i\omega
    \end{pmatrix}\\
    =& U(\sigma_{x}-\mathbb{I})+V(\sigma_{z}+i\sigma_{y})
    \end{aligned}
\end{equation}

where $\sigma_{x,y,z}$ are the Pauli matrices and $U=(\lambda+i\omega)$, $V=\dfrac{\lambda_{2}-\lambda_{1}}{2}$. Now exponentiating the $\mathcal{W}$ matrix we get,

\begin{equation}
          e^{\mathcal{W}t}=e^{-Ut}\bigg(\mathbb{I}Cosh(Ut)+\bigg(\sigma_{x}+\dfrac{V}{U}(\sigma_{z}+i\sigma_{y})\bigg)Sinh(Ut)\bigg)
 \end{equation}

  which yields the time dynamics of the conditional probabilities which are now complex~\cite{ComplexProb1,ComplexProb2} as a consequence of complex transition rates. The individual conditional probabilities, although complex, obey conservation and take the following form,

\begin{equation}
    \begin{aligned}
        &P(1,t)=\dfrac{1+V/U}{2}+\dfrac{(1-V/U)a-(1+V/U)b}{2}e^{-2(\lambda+i\omega) t}\\
         &P(2,t)=\dfrac{1-V/U}{2}+\dfrac{(1+V/U)b-(1-V/U)a}{2}e^{-2(\lambda+i\omega) t}.\\
    \end{aligned}
\end{equation}

\begin{figure}[ht!]
     \begin{center}
\includegraphics[height=7cm]{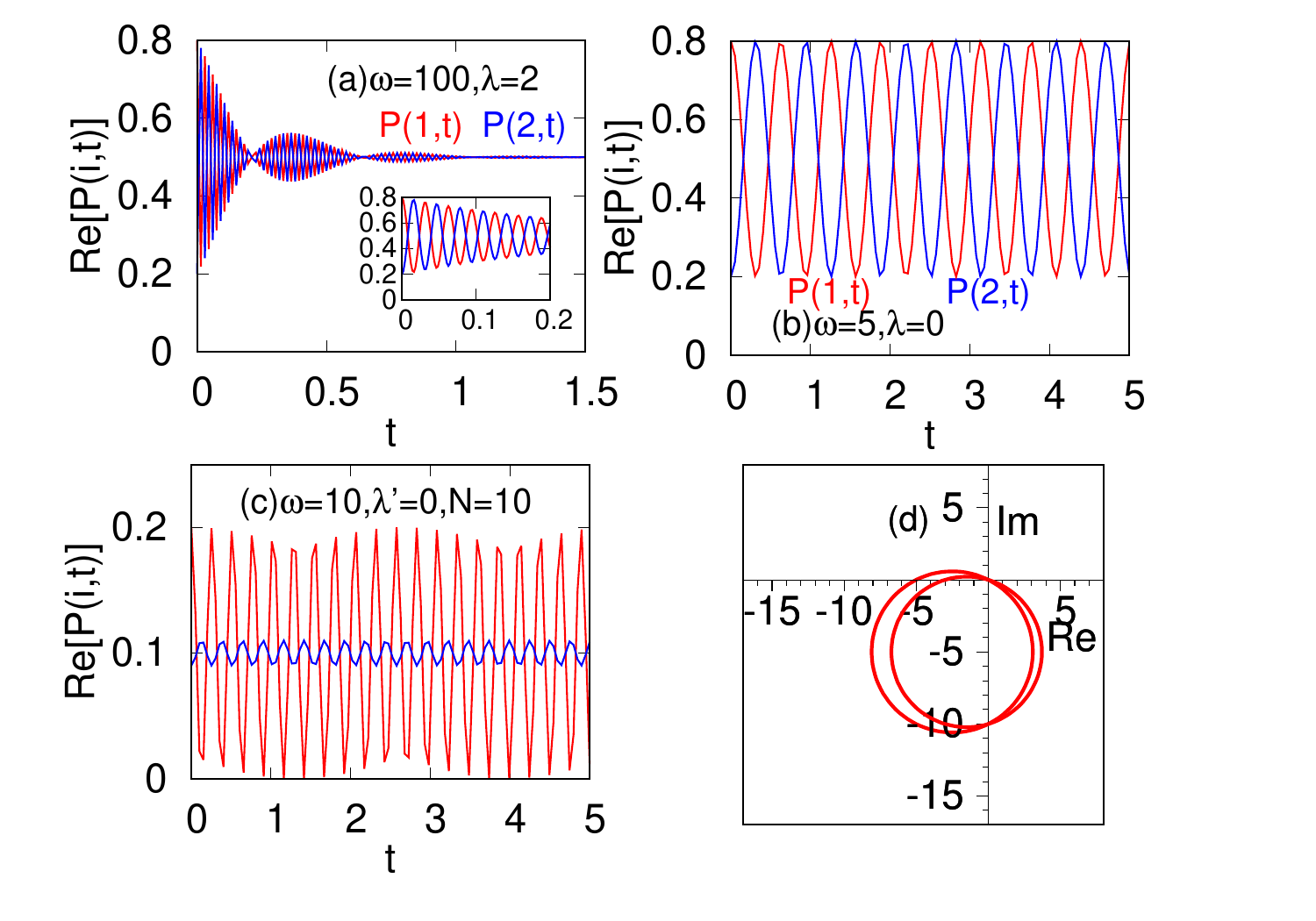}
\end{center}
   \caption{Panel (a) shows oscillations in the conditional probabilities in the pre-thermal regime for complex transition rates with $\omega=100$ and $\lambda_{1}=2.5, \lambda_{2}=1.5$ for a dichotomous process. Inset shows oscillations on a smaller time scale. Panel (b) shows persistent long time oscillations for imaginary transition rates with $\omega=5$ for a dichotomous process. Panel (c) shows oscillations for two specific states having initial probabilities $0.2$ (red) and $0.09$ (blue) for a N=10 state system with imaginary transition rates with $\omega=10$. Real part of the complex probabilities have been plotted. Panel (d) shows the Gershgorin discs in case of a dichotomous process with transition rates $\lambda_{1}+i\omega=1.5+i5$ and $\lambda_{2}+i\omega=2.5+i5$. The eigenvalues are $0,-2(\lambda+i\omega)$, where $\lambda$ is the mean transition rate.  }{\label{fig2}}
  % \label{fig:subfigures}
\end{figure}

These equations show that the complex conditional probabilities show pre-thermal oscillations, the range of which can be altered by tuning $\lambda$ suitably. The oscillatory behaviour is driven by the imaginary part of the $\mathcal{W}$ matrix which means there is a back and forth transition between the two states. However, the overall relaxation is guided by the real part of $\mathcal{W}$ matrix. In the case of $\lambda=0$, persistent long time oscillations in the probabilities are observed, with no signature of equilibriation. Therefore, an imaginary $\mathcal{W}$ is truly a non-equilibrium scenario where the initial state we started with returns again and again to the same state without damping. There is a specific initial state with equal weights for both states, which does not show any oscillatory dynamics and remain pinned to the values$\dfrac{1\pm V/U}{2}$. Fig~\ref{fig2}(a) show robust pre-thermal oscillations for complex transition rates with parameters $\omega=100,\lambda_{1}=2.5,\lambda_{2}=1.5$, which eventually decay with time as the system equilibriates. Panel (b) shows long time persistent oscillations corresponding to purely imaginary transition rates with $\omega=5$ which is clearly a non-equilibrium phenomena. Panel (d) shows the Gershgorin circles for the case of complex transition rates with $\omega=5, \lambda_{1}=2.5,\lambda_{2}=1.5$. Here the union of the circles contain the eigenvalues $0,-2(\lambda+i\omega)$.

{ Stochasticity enters through the real part of the transition rate, $\lambda$, which is driven by thermal fluctuations due to coupling with a bath while the imaginary part, $\omega$, is deterministic in nature originating from bounce back or reflection mechanisms intrinsic to the system. These mechanisms induce persistent oscillations between the states. Increasing the temperature of the thermal bath increases $\lambda$, leading to faster relaxation dynamics and a shorter pre-thermal oscillatory regime. While $\lambda$ is dissipative in nature, $\omega$ is oscillatory in nature and induces unitary-like dynamics in conditional probabilities. Therefore, stochasticity is not a pre-requisite for observing persistent oscillations. For example, in Rydberg atom systems~\cite{Rydberg1,Rydberg2}, laser-driven transitions may include reflections from impurity states, effectively introducing an imaginary component to the transition rate. Such a rebound mechanism prevents equilibration and sustains oscillations. Also, in models like the PXP model~\cite{Scar1,Scar2}, where kinematic constraints are responsible for fidelity oscillations from specific initial states, these constraints can be mapped to an effective imaginary transition rate in the master equation framework. This allows the modelling of such non-dissipative periodic revivals.    }

We next expand our state space to $N$ dimensions. We choose the $\mathcal{W}$ matrix to be symmetric with off-diagonal elements $W_{jk}={\lambda'}+i\omega, j\neq k$ and the diagonal elements $W_{kk}=-(N-1)({\lambda'}+i\omega)$. Then,

\begin{equation}
    \mathcal{W}=-(N-1)({\lambda'}+i\omega) \mathbb{I}+ ({\lambda'}+i\omega)(\mathcal{X}-\mathbb{I}). 
\end{equation}

Here, $\mathcal{X}_{ij}=1$ for all $i,j$. Then,

\begin{equation}
\begin{aligned}
   e^{\mathcal{W}t}=&e^{-N({\lambda'}+i\omega)t}\bigg[\dfrac{\mathcal{X}}{N}\sum_{n=1}^{\infty}\dfrac{(N({\lambda'}+i\omega)t)^{n}}{n!}+\mathbb{I}\bigg]\\=&\bigg(\mathbb{I}-\dfrac{\mathcal{X}}{N}\bigg)e^{-N({\lambda'}+i\omega)t}+\dfrac{\mathcal{X}}{N}
 \end{aligned}
 \end{equation}

In the previous equation we have used $\mathcal{X}^{n}=N^{n-1}\mathcal{X}$. Now, let us take an initial state $\{P_{i}\}$, $i$=$1,2,3$...$N$. Then,

\begin{equation}
    P(i,t)=\bigg(P_{i}-\dfrac{1}{N}\bigg)e^{-N({\lambda'}+i\omega)t}+\dfrac{1}{N}
\end{equation}

For ${\lambda'}=0$, the dynamics again correspond to an oscillatory non-equilibrium dynamics with the only exception of the initial state for which $P_{i}=1/N \forall i$. We see that we have to impose a constraint on initial states if the real and imaginary parts of the probabilities have to be positive. Only such initial states would correspond to physical situations for which $P_{i}\leq 2/N \forall i$. Fig~\ref{fig2}(c) shows persistent oscillations for $N=10$ when ${\lambda'}=0$ for two example states for which initial probabilities were taken to be $0.2$ and $0.09$. Further, it is understandable that for ${\lambda'} \neq 0$, the system will equilibriate (after pre-thermal oscillations) to the thermal value $1/N$ which clearly depends on dimensionality.

\subsection{Violation of probability conservation}

\begin{figure}[ht!]
     \begin{center}
\includegraphics[height=6cm,trim={0 0 6cm 0},clip]{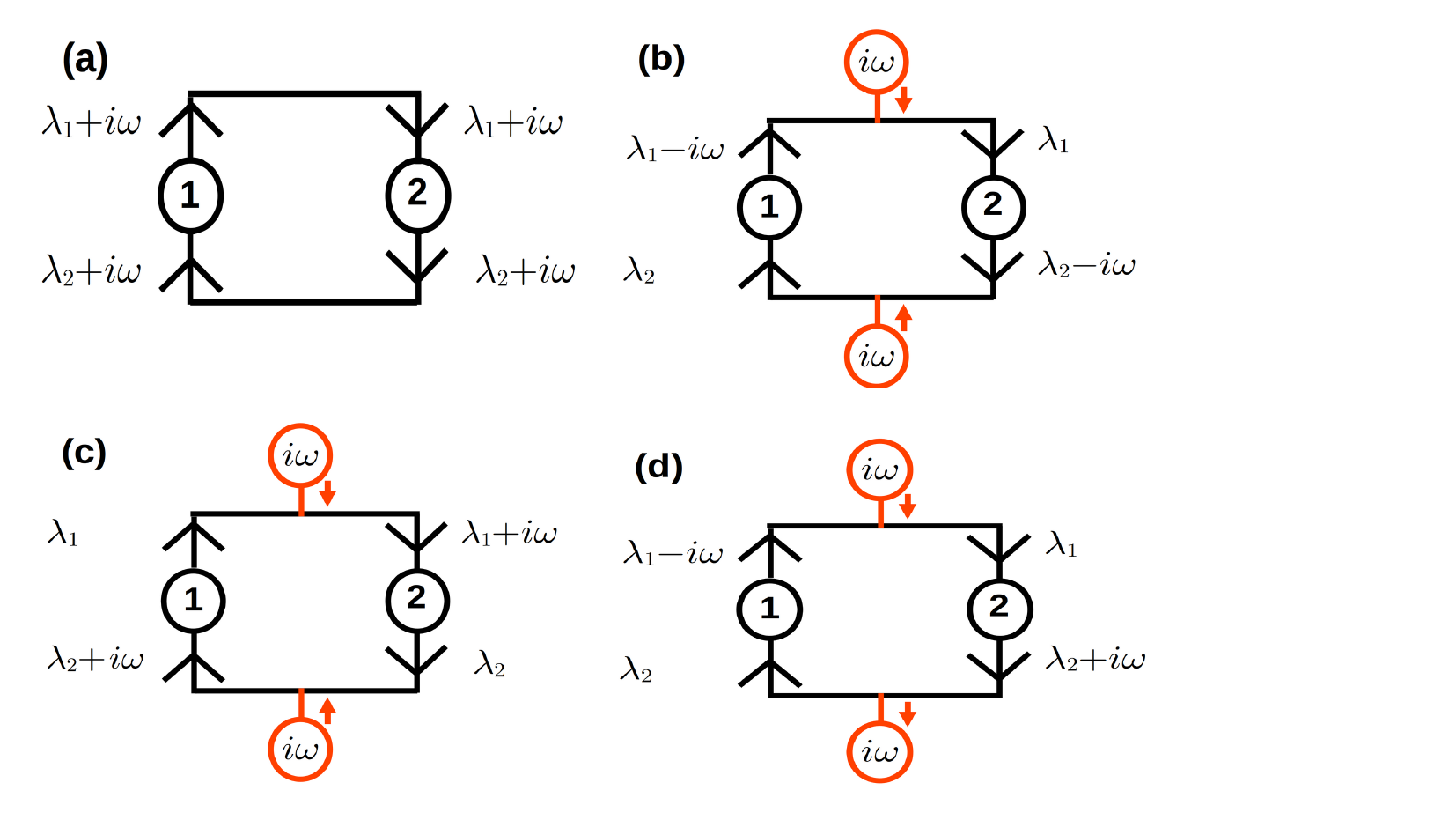}
\end{center}
   \caption{Panel (a) shows rate of probability exchange between two states in a dichotomous process when the transition rates are complex. Panels (b) and (c) show that probability has to pumped ``in" to both the arms to maintain the given transition rates via an external drive. Panel (d) shows that unlike the two previous cases, probability needs to be pumped ``in" from one arm and pumped ``out" from the other via an external drive to maintain the given transition rates. \label{fig3} }
  % \label{fig:subfigures}
\end{figure}

We also explore cases where the sum of the elements of the columns of the $\mathcal{W}$ matrix is non-zero. We again look at the dichotomous process but now we perturb the system with a drive of angular frequency $\omega$. If we drive both the states with same frequency simultaneously, it corresponds to a scenario where we couple the original relaxation matrix with an identity matrix with $i\omega$ as diagonal entries. This means with the regular loss term , we have a term which also returns the state to itself. Schematic diagram Fig~\ref{fig3}(b) depicts the possible scenario where probability has to be pumped ``in" both arms from an external source as opposed to panel (a) where we have shown the same for complex transition rates as discussed in the previous section. The probabilities remain the same as that of the undriven case modulo an oscillating phase factor in time as given by,
\begin{equation}
\begin{aligned}
   & P(1,t)=e^{i\omega t}\bigg(a-\dfrac{\lambda_{2}b-\lambda_{1}a}{2\lambda}(e^{-2\lambda t}-1)\bigg)\\
    &  P(2,t)=e^{i\omega t}\bigg(b-\dfrac{\lambda_{1}a-\lambda_{2}b}{2\lambda}(e^{-2\lambda t}-1)\bigg).
      \end{aligned}
\end{equation}

The probabilities also remain conserved modulo an oscillating phase factor in time. Similarly, we can couple the relaxation matrix with the Pauli matrix $\sigma_{x}$ multiplied with $i\omega$. Schematic diagram Fig~\ref{fig3}(c) depicts the possible scenario where probability has to be again pumped ``in" both the arms from an external source similar to the previous case. The conditional probabilities remain the same as that for the complex transition rates modulo an oscillating phase in time. This is because the relaxation matrix can be reframed as the $\mathcal{W}$ for the complex transition rate plus the identity matrix multiplied with $i\omega$. Like the previous case the total probability remains conserved modulo a phase factor oscillating in time.

However, the more interesting case arises when we rotate the complex probability vector for one state in the anti-clockwise sense and the other in the clockwise sense. This induces an instability in the system and the system eventually disintegrates/melts and the individual probabilities go to zero. Schematic diagram Fig~\ref{fig3}(d) depicts the possible scenario where probability has to be pumped ``in" one of the arms and pumped ``out" from the other from an external source as opposed to the previous two cases. In this case the $\mathcal{W}$ matrix is,

\begin{figure}[ht!]
     \begin{center}
\includegraphics[height=3.5cm]{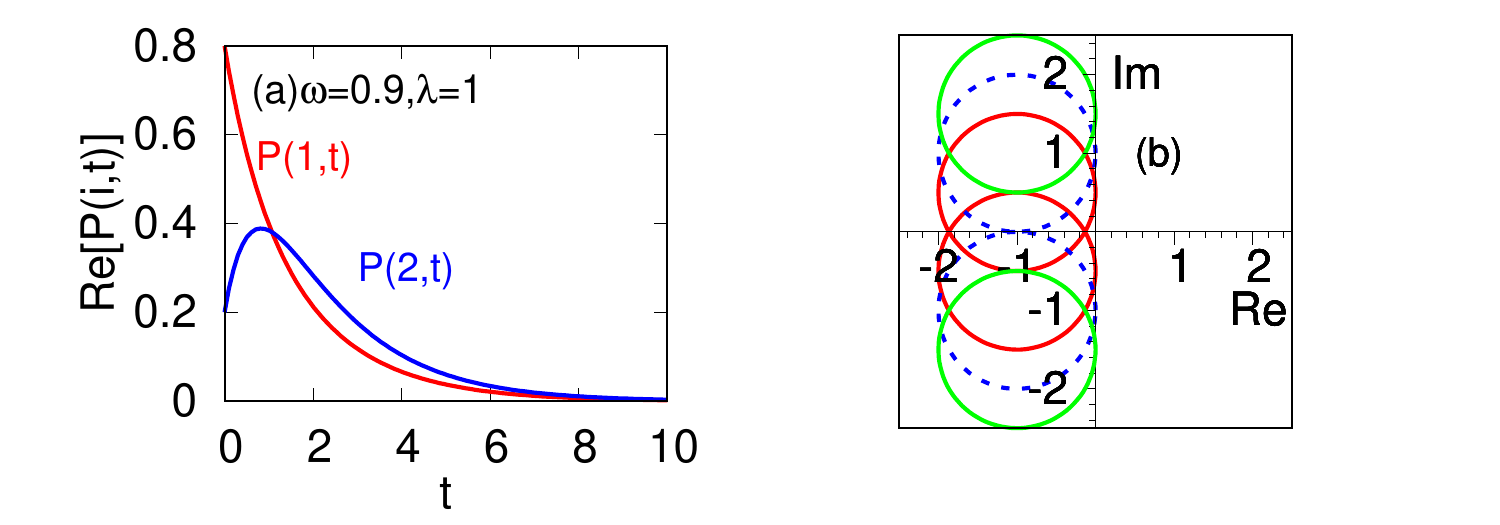}
\end{center}
   \caption{Panel (a) shows that both $Re[P(1,t)]$ and $Re[P(2,t)]$ decay to zero which means the system eventually becomes unstable to the drive and disintegrates/melts. Here, $\omega=0.9$  and $\lambda=1$ with $\lambda_{1}=\lambda_{2}$. Panel (b) shows the Gershgorin disc pairs for three cases: $\omega > \lambda$ when the circles are disjoint and the corresponding eigenvalues are complex (green), $\omega=\lambda$  when the circles just touch at value $-\lambda$ on the real axis (blue) and when $\omega<\lambda$ where there are two distinct real eigenvalues (red).}{\label{fig4}}
  % \label{fig:subfigures}
\end{figure}

\begin{equation}\begin{aligned}
     &\mathcal{W}=\begin{pmatrix}
        -\lambda_{1}+i\omega & \lambda_{2}\\
        \lambda_{1} &  -\lambda_{2}-i\omega   \end{pmatrix}\\
        &=\lambda(\sigma_{x}-\mathbb{I})+\dfrac{\lambda_{2}-\lambda_{1}}{2}(\sigma_{z}+i\sigma_{y})+i\omega\sigma_z.
\end{aligned}
\end{equation}

Here, $\lambda=(\lambda_{1}+\lambda_{2})/2$ is the mean transition rate. Then,

\begin{equation}
    e^{\mathcal{W}t}=e^{-\lambda t}\bigg[\mathbb{I}Cosh(Yt)+\dfrac{(\mathcal{W}+\lambda \mathbb{I})}{Y} Sinh(Yt)\bigg]
\end{equation}

where, $Y=\sqrt{\lambda^{2}-\omega^{2}+i\omega(\lambda_{2}-\lambda_{1})}$.Let us focus on the specific case of the symmetric dichotomous process where $\lambda_{1}=\lambda_{2}$ such that now $Y=\sqrt{\lambda^{2}-\omega^{2}}$. Then the conditional probabilities take the following form,

\begin{equation}
    \begin{aligned}
        P(1,t)=&e^{-\lambda t}\bigg[a Cosh(Yt)+\dfrac{\lambda b+i\omega a}{Y}Sinh(Yt)\bigg]\\
         P(2,t)=&e^{-\lambda t}\bigg[b Cosh(Yt)+\dfrac{\lambda a-i\omega b}{Y}Sinh(Yt)\bigg].
         \end{aligned}
\end{equation}

We look at the case where $\lambda>\omega$ and find that both probabilities decay to zero with time implying that an instability has been induced by the drive. In other words, the length of the probability vector damps to zero with time. Fig~\ref{fig4} (a) shows the probability damping in this case for a suitable choice of parameters as specified in the caption to the figure. Also shown in panel (b) are the Gershgorin circle pairs for three cases. For $\omega>\lambda$ the circles are disjoint and there are no real eigenvalues. For $\omega=\lambda$ the circles touch the real axis at $-\lambda$ and upon decreasing $\omega$ further the circles overlap, touching the real axis at two points $-\lambda \pm \sqrt{\lambda^{2}-\omega^{2}}$ which are the eigenvalues of the $\mathcal{W}$ matrix. So the gap between the circles close and they merge as we tune the value of $\omega$ with respect to $\lambda$ or vice-versa. It is worthwhile to mention that for $\omega>\lambda$, similar dissipation will be observed in the length of the probability vector accompanied with pre-disintegration oscillatory dynamics in length.  

\section{Conclusion}

In summary, we have extended the formalism of the classical master equation, firstly, to complex transition rates which induce pre-thermal oscillations in the conditional probabilities. Further, for purely imaginary transition rates, the system does not equilibrate. Instead it shows persistent oscillations in the probability dynamics. We see this both for a two state system as well as an $N$-state system and find the feature to be quite generic. Secondly, we look at a dichotomous stochastic processes in which probability is being pumped ``in" and ``out" of the system from outside to maintain the transition rates of the system. We call such external perturbations, external drive. In such a scenario, we looked at cases where the total probability is conserved modulo some oscillating phase in time. However, a more interesting case arises when the complex probability phasor was rotated clockwise for one state and anti-clockwise for the other state through external drive. In such a case the ultimate fate of the system is disintegration/melting where two distinct states cease to exist and the individual conditional probabilities go to zero. Thus the system becomes unstable in response to the external drive. 

These novel oscillatory and dissipative dynamics can be experimentally probed in ultracold platforms with two level systems such as Rydberg atoms~\cite{Rydberg1,Rydberg2}. The underlying physics of these phenomena can be used to understand broader questions regarding oscillatory dynamics in the fidelity oscillations of some special states in isolated quantum systems. It is not very difficult to extend the discussion to quantum systems since in the quantum master equation, the transition rates given by the Fermi-Golden rule~\cite{FermiGolden} can be similarly coupled with an imaginary parameter to explore dynamics in such systems. 

{Moreover, since noise has been shown to contribute constructively in phenomena like  NES~\cite{NES1,NES2,NES3}, long-lived entanglement in noisy qubits~\cite{npj}, and stochastic resonance~\cite{sr}, as a future direction, it can be explicitly included in our model to study emergent novel phenomena. Incorporating such effects into our framework would broaden the applicability to real-world systems, especially those described by non-Hermitian or driven-dissipative dynamics. Given that noise can stabilize non-equilibrium states, the interplay of noise and scars ~\cite{scar_noise} in quantum systems is indeed a compelling direction. }

\end{document}